\documentclass[conference]{IEEEtran}
\IEEEoverridecommandlockouts
\usepackage{cite}
\usepackage{amsmath,amssymb,amsfonts}
\usepackage{algorithmic}
\usepackage{graphicx}
\usepackage{textcomp}
\usepackage{xcolor}

\usepackage{multirow}
\usepackage{framed}
\usepackage{multicol}
\usepackage{booktabs}

\usepackage{listings}

\usepackage{changepage}

\usepackage{dblfloatfix}

\usepackage{hyperref}

\usepackage{balance}

\newcommand{\mcode}[1]{$\tt #1$} 

\definecolor{lightgray}{rgb}{0.83, 0.83, 0.83}
\definecolor{formalshade}{rgb}{0.95,0.95,1}
\definecolor{darkblue}{rgb}{0.0, 0.0, 0.55}

\def\BibTeX{{\rm B\kern-.05em{\sc i\kern-.025em b}\kern-.08em
    T\kern-.1667em\lower.7ex\hbox{E}\kern-.125emX}}
\begin{document}

\title{REST vs GraphQL: A Controlled Experiment}

% %\title{Do Clients need GraphQL APIs? \\  A mixed-method study}

\author{\IEEEauthorblockN{Gleison Brito\IEEEauthorrefmark{1}, Marco Tulio Valente\IEEEauthorrefmark{1}}
\IEEEauthorblockA{\IEEEauthorrefmark{1}ASERG Group, Department of Computer Science (DCC), Federal University of Minas Gerais, Brazil\\
\{gleison.brito, mtov\}@dcc.ufmg.br}
}

% \author{\IEEEauthorblockN{Anonymous}
% \IEEEauthorblockA{ \\
%  }
% }

\maketitle

\begin{abstract}
GraphQL is a novel query language for implementing service-based software architectures. The language is gaining momentum and it is now used by major software companies, such as Facebook and GitHub. However, we still lack empirical evidence on the real gains achieved by GraphQL, particularly in terms of the effort required  to implement queries in this language. Therefore, in this paper we describe a controlled experiment with 22 students (10 undergraduate and 12 graduate), who were asked to implement eight queries for accessing a web service, using GraphQL and REST. Our results show that GraphQL requires less effort to
implement remote service queries, when compared to REST
(9 vs 6 minutes, median times). 
These gains increase when REST queries include
more complex endpoints, with several parameters.
Interestingly, GraphQL outperforms REST even
among more experienced participants (as is the
case of graduate students) and among participants
with previous experience in REST, but no previous
experience in GraphQL. 

\end{abstract}

%\begin{IEEEkeywords}
%component, formatting, style, styling, %insert
%\end{IEEEkeywords}

\section{Introduction}\label{sec:intro}
 
GraphQL is a query language for implementing web service architectures~\cite{graphql2015}. The language was internally developed at Facebook, as a solution to several problems faced by them when using standard architectural styles, such as REST. In 2015, Facebook open-sourced the definition and implementation of GraphQL. As a result, the language started to gain  momentum and it is now supported by major Web APIs, including the ones provided by GitHub, Airbnb, Netflix, and Twitter. In December 2018, Facebook decided to transfer GraphQL to a non-profit organization, called GraphQL Foundation.

GraphQL is as an alternative to REST-based applications~\cite{fieldingT02,fieldingT00,fielding:2000}. 
To understand GraphQL's differences from REST, we must remember that endpoints are the key abstraction provided by REST. In REST, an endpoint is defined by an URL and a list of parameters. For example, in GitHub's REST API
\begin{lstlisting}[numbers=none]
GET /search/repositories?q=stars:>100
\end{lstlisting}

\noindent is an endpoint that returns data about GitHub repositories with more than 100 stars. Since REST endpoints rely on HTTP resources to support queries (URLs, GET/PUT parameters, etc), they can be considered as low-level abstractions. 
By contrast, GraphQL is a full data query language to implement web-based services, centered on high-level abstractions, such as schemas, types,
queries, and mutations. For example, the previous REST query is implemented in GraphQL as follows:
\begin{lstlisting}
query searchRepos {
 search(query:"stars:>100", first:100, type:REPOSITORY){
    nodes{
      ... on Repository{ nameWithOwner }
    }
  }
}
\end{lstlisting}

When using GraphQL, clients can define exactly the data they require from service providers. In our previous REST example, the server returns
a JSON document with 94 fields, although the client only consumes one field (the repository's name). This problem is called \textit{over-fetching}~\cite{schrock2015,buna2017}. On the other
hand, in GraphQL, clients can precisely specify the fields they require from servers (in our example, just \mcode{nameWithOwner}, line 4).

Previous studies compared REST and GraphQL, but
mostly under a quantitative perspective. For example, Brito et al.~\cite{brito2019} investigated the  gains of migrating to GraphQL queries performed by seven GitHub API clients. Wittern et al.~\cite{wittern2018} performed a study to evaluate the gains achieved with a tool that automatically generates GraphQL wrappers from REST APIs. However, to our knowledge, \textbf{we still lack studies that contrast the effort and the perceptions of developers when implementing remote queries using REST and GraphQL}.
As a contribution to close this gap, in this paper we present the results of a controlled experiment where we asked 22 students to implement a set of queries for accessing GitHub services. We 
anchored the experiment on this particular service because
GitHub supports a REST implementation and also a GraphQL-based version. Therefore, 
we instructed the students to implement half of the proposed
queries in REST and the other half in GraphQL.

We ask two questions in this paper:

\vspace{4pt}
\noindent \textit{RQ1: How much time do developers spend when implementing queries in REST and GraphQL?} Our intention is to investigate possible gains achieved by GraphQL; not in terms of transferring less data to clients, i.e., avoiding over-fetching, but on demanding less effort to implement the queries.
To provide a more solid answer, we expanded this first RQ in three related sub-questions: ({\em RQ1.1}) How does this time vary between the types of queries? ({\em RQ1.2}) How does this time vary among undergraduate and graduate students? and ({\em RQ1.3}) How does this time vary depending on the participants' experience in REST and GraphQL?

\vspace{4pt}
\noindent \textit{RQ2: What are the participants’ perceptions about REST and GraphQL?} With this second question, our intention is to provide qualitative data on the experience of the experiment's participants when implementing the proposed GitHub queries. Basically, we surveyed the participants about their perceptions on GraphQL, REST, and our experiment, in general.
\vspace{4pt}

In summary, our results show that GraphQL requires less effort to
implement service queries when compared to REST
(9 vs 6 minutes, median times). However, we found that the gains are mostly restricted to REST queries that have several parameters.
Interestingly, GraphQL outperforms REST even
among experienced participants (as is the
case of graduate students) and among participants
with previous experience in REST and no previous
experience in GraphQL. Finally, when surveyed,
the participants mentioned two key benefits of
using GraphQL: (1) tool support for building
and testing the queries (particularly the support
provided by auto-complete features); (2) a
syntax and semantics closed to standard
programming languages, centered on concepts
such as schemas, types, queries, interfaces, and objects. 

The rest of this paper contains seven sections. Section~\ref{sec:background} provides a brief introduction to REST and GraphQL using GitHub APIs as example. Section~\ref{sec:design} describes the design of the proposed experiment. Section~\ref{sec:results} presents the results of the two proposed research questions. 
Section~\ref{sec:discution} discuss the main findings of the experiment.
Threats to validity are discussed in Section~\ref{sec:threats}; and related work is discussed in Section~\ref{sec:relWork}. Finally, Section~\ref{sec:conclusion} concludes the paper.

\section{Background}\label{sec:background}
% This section presents some concepts necessary for understanding GraphQL. GraphQL APIs implement a \emph{schema}. In this \emph{schema} the types and relationships between the data exposed by the API are defined.
% While REST APIs expose data across multiple endpoints, GraphQL APIs use a graph. The \emph{schema}  is a representation of the graph. The GraphQL schema that defines the types and relations of exposed data, including operations to query or mutate data. The schema defines the API data as object types, that contains fields and each field has a name and also a type.

This section presents a short overview of Web services, and the architectural models REST and GraphQL. The most important characteristics of each architectural model are presented for a better understanding of the motivations behind our study. For a detailed presentation of GraphQL, we refer the reader to its documentation~\cite{graphql2015}. For REST, we recommend the doctoral thesis that introduced this concept~\cite{fielding:2000}.

\subsection{Web Services}

Web services are collections of protocols and standards used to exchange data between web systems. 
Software applications written in multiple programming languages and running on various platforms can use web services to exchange data on computer networks, such as the Internet. 
These services provide interoperability between systems communication~\cite{mizouni2011}.
There have been several implementations that provide solutions for this concept, e.g., SOAP, REST, and GraphQL.

\subsection{REpresentational State Transfer (REST)}

REST is an architectural style for implementing distributed systems. The style defines a set of constraints intended to improve performance,  availability, and scalability and it is based on a traditional client-server paradigm~\cite{fieldingT02, fieldingT00, fielding:2000}. REST-based APIs are the ones that follow the constraints defined by the REST style. REST also defines a uniform interface for system components based on resource identification and dynamic data provision. In REST-based APIs, data is exposed by means of \textit{endpoints}. Each endpoint returns data about one resource and each resource has a predefined set of fields.

For example, GitHub's REST API provides 366 endpoints. An example of an endpoint is
\begin{lstlisting}[numbers=none]
GET /users/torvalds/repos
\end{lstlisting}

This endpoint returns the list of public repositories of a given user, e.g., \textit{torvalds}. The following listing shows a fragment of the returned JSON. It contains 93 fields, e.g., \mcode{full\_name} (line 3), \mcode{owner} (line 5--8), \mcode{created\_at} (line 10), among others. 
\begin{lstlisting}
[
  {
    "full_name": "torvalds/libdc-for-dirk",
    "private": false,
    "owner": {
      "login": "torvalds",
      ...
    },
    "created_at": "2017-01-17T00:25:49Z",
    ...
  },
  ...
]
\end{lstlisting}

\subsection{GraphQL}
 
In GraphQL, service data is exposed as a graph~\cite{hartig2017}, defined by means of a schema. Each node of this graph/schema represents objects and contains fields. Each field has a name and a type. Edges appear when a field references another object. Clients access a GraphQL service through a single endpoint, which is used to submit queries.

GraphQL provides a domain specific language for defining schemas, including types and queries. For example, GitHub's GraphQL API has a schema with types such as \textit{Repositories} and \textit{Users}, among other entities.\footnote{Available at \url{https://github.com/octokit/graphql-schema}} The following listing shows a fragment of \mcode{Repository} and \mcode{Language} types.
\begin{lstlisting}
interface Node {
  id: ID!
}

type Repository implements Node {
    nameWithOwner: String!
    primaryLanguage: Language!
    ...
}

type Language implements Node {
    id: ID!
    name: String!
    color: String
}
\end{lstlisting}

Like many type systems, GraphQL also supports \textit{interfaces}. An interface is an abstract type that includes fields that a type must define when implementing the interface. Most types from GitHub's schema---including \mcode{Repository} and \mcode{Language}---implements the \mcode{Node} interface (lines 1--3). This interface has only one field, called \mcode{id}, that represents a unique identifier. The \mcode{Repository} type contains 71 fields. However, to the sake of clarity, our example only shows two fields: \mcode{nameWithOwner} (line 7) and \mcode{primaryLanguage} (line 8). The \mcode{primaryLanguage} field is of type \mcode{Language} (lines 12--16), which contains three fields: \mcode{id}, \mcode{name}, and \mcode{color} (which is the color defined for the language on GitHub's web interface). The \mcode{!} symbol means that a field value must not be \mcode{null}. 

In GraphQL schemas, queries are defined using a special type, called \mcode{Query}. The following listing shows a fragment of a \mcode{Query} type, with only one query, called \mcode{repository} (line 3), which has two parameters: \mcode{name} and owner. Both parameters are of type \mcode{String}. This query returns an object of a \mcode{Repository} type.%\\[-0.85cm]
\begin{lstlisting}
type Query {
  repository(name: String!, owner: String!): Repository
  ...
}
\end{lstlisting}

Finally, GraphQL defines a query language, used by clients to perform queries. The following listing shows a example of  \mcode{repository} query. The query (\mcode{exampleRepository01}) returns the full name (\mcode{nameWithOwner}, line 3) of \textsc{facebook/react}. %\\[-0.48cm]
\begin{lstlisting}[] 
query exampleRepository01{
  repository(owner:"facebook", name:"react"){
    nameWithOwner
  }
}
\end{lstlisting}

The result of \mcode{exampleRepository01} query is presented in the following listing. As we can see, the result is a JSON object resembling the structure of the query. 
\begin{lstlisting}
{
  "data": {
    "repository": {
      "nameWithOwner": "facebook/react"
    }
  }
}
\end{lstlisting}

\subsection{Differences between REST and GraphQL}
To conclude, we summarize the key differences between REST and GraphQL. In GraphQL, service data is exposed as a graph, represented by a schema. By contrast, in REST, server applications implement a list of endpoints.
GraphQL also defines a query language that allows clients to specify precisely the fields they demand from servers. Furthermore, in REST services, the queries are defined by means of endpoints. Each endpoint returns a predefined set of fields that represents data about some resource. On the other hand, in GraphQL, the response resembles the query structure.

\section{Experiment Design}\label{sec:design}

In this paper, we describe a controlled experiment to compare two technologies for implementing web services: REST and GraphQL.
We aim to reveal which technology requires less effort to implement queries to Web services.
We ask the following research questions:

\begin{itemize}
    \item \noindent RQ1: \textit{How much time do developers spend implementing queries in REST and GraphQL?} In fact, to provide an in-depth understanding of this first question, we investigate three related questions:
    \begin{itemize}
        \item  \noindent RQ1.1: \textit{How does this time vary among the types of queries?}
        %\item  \noindent \textbf{RQ1.2}: How does this time vary between the tasks?
        \item  \noindent RQ1.2: \textit{How does this time vary among undergraduate and graduate students?}
        \item  \noindent RQ1.3: \textit{How does this time vary depending on the participants' experience in REST and GraphQL?}
    \end{itemize}
    \item  \noindent RQ2: \textit{What are the participants' perceptions about REST and GraphQL?} With this second RQ, our goal is to provide qualitative data about the implementation of API queries using REST and GraphQL, based on the perceptions and views of the participants.
\end{itemize}

\begin{table}[t!]
\caption{Experiment Tasks}
\begin{tabular}{@{}c@{\hspace*{2pt}}c@{\hspace*{2pt}}p{7cm}@{}}
\toprule
\multicolumn{1}{c}{\textbf{Type}} & \textbf{Task} & \multicolumn{1}{c}{\textbf{Description}}                                                                                                                                                                      \\ \midrule
\multirow{9}{*}{\begin{tabular}[c]{@{}c@{}}Search \\ Repositories\end{tabular}}           & \multirow{3}{*}{T1}            & Implement a query that returns the full name (\textsc{owner/name}) and the description of the top-10 most starred Python repositories, sorted in descending order.                              \\ \cmidrule{2-3} 
                                  & \multirow{3}{*}{T2}            & Implement a query that returns the number of stars and the number of forks of the top-10 most starred repositories, sorted in descending order.                                                                            \\ \cmidrule{2-3} 
                                  & \multirow{3}{*}{T3}            & Implement a query that returns the URL and owner login of the top-10 most starred Java repositories, created after Jan-01-2018, sorted in descending order.                                                       \\ \midrule
\multirow{4}{*}{\begin{tabular}[c]{@{}c@{}}Search \\ Users\end{tabular}}          & \multirow{2}{*}{T4}            & Implement a query that returns the URL of 10 users with more than 10,000 followers, sorted in descending order.                                                                                                   \\ \cmidrule{2-3} 
                                  & \multirow{2}{*}{T5}            & Implement a query that returns the login of 10 individual users (i.e., non-organizations) with more than 10,000 repositories.                                                                                                                      \\ \midrule
\multirow{2}{*}{Repository}       & \multirow{2}{*}{T6}            & Implement a query that returns the primary language, the description, and URL of \textsc{facebook/graphql} repository.                                                                            \\ \midrule
\multirow{5}{*}{User}             & \multirow{2}{*}{T7}            & Implement a query that returns the number of followers and the number of repositories of the user \textsc{torvalds}.                                                                             \\ \cmidrule{2-3} 
                                  & \multirow{3}{*}{T8}            & Implement a query that returns the number of watchers and the number of stars of the first 10 repositories owned by \textsc{facebook}, sorted by creation date, in descending order. \\ \bottomrule
\end{tabular}
\label{tab:tasks}
\end{table}

Before starting the controlled experiment, we performed a pilot study with two graduate students. Both participants had previous experience only in REST; for this reason, the first author of this paper presented a short talk on GraphQL (one hour).  They implemented eight queries, one student using REST and the other student using GraphQL. We used their comments and observations to calibrate our study.

\subsection{Tasks}
To answer the proposed research questions, we rely on a controlled experiment involving four types of queries to GitHub: \emph{search repositories}, \emph{search users}, \emph{repository}, and \emph{user}.
We selected these endpoints due to their relevance. For example, in order to study the gains achieved by GraphQL due to the lack of over-fetching, Brito et al.~\cite{brito2019} implemented 14 queries used in seven recent empirical software engineering papers. These queries use exactly the same endpoints we selected for our experiment.

After selecting the endpoints, we elaborated three tasks requiring the implementation of \emph{search repositories} queries, two requiring \emph{search users} queries, 
one requiring a \emph{repository} query, and two demanding \emph{user} queries. These eight queries are described in Table~\ref{tab:tasks}.
\emph{Search repositories} and \emph{search users} are generic queries that return data about repositories and users, respectively, using parameters to filter the results. \textit{Repository} and \textit{User} queries are specific queries that return data about only one repository or user, respectively.

\subsection{Subjects Selection}

We performed our controlled study with 22 subjects, including 10 undergraduate students and 12 graduate students. All subjects have at least one year of programming experience. Additionally, as we can see in Table~\ref{tab:prev_knowledge}, 
11 subjects have previous experience with REST, and four have experience with both REST and GraphQL. We also have seven subjects without experience in any of these technologies. It is also worth noting that no participants have had experience only in GraphQL.

\begin{table}[h!]
\centering
\caption{Subjects experience in REST and GraphQL}
\label{tab:prev_knowledge}
\begin{tabular}{ c c c c }
\toprule
\bf{REST} & \bf{GraphQL} & \bf{REST and GraphQL} & \bf{None} \\ 
\midrule
 11 (50\%) 	& 0 (0\%) 	& 4 (18.2\%) & 7 (31.8\%) \\
\bottomrule
\end{tabular}
\end{table}

% \usepackage{multirow}
\begin{table}[t!]
\centering
\caption{Tasks allocation among participants (Groups A and B) and between treatments (REST and GraphQL)}
\begin{tabular}{cll}
\toprule
\multirow{2}{*}{\textbf{Tasks}} & \multicolumn{2}{c}{\textbf{Group}}                                   \\ \cmidrule{2-3} 
                                & \multicolumn{1}{c}{\textbf{A}} & \multicolumn{1}{c}{\textbf{B}} \\ \midrule
\multicolumn{1}{c|}{T1}         & REST                           & GraphQL                        \\
\multicolumn{1}{c|}{T2}         & GraphQL                        & REST                           \\
\multicolumn{1}{c|}{T3}         & REST                           & GraphQL                        \\
\multicolumn{1}{c|}{T4}         & GraphQL                        & REST                           \\
\multicolumn{1}{c|}{T5}         & REST                           & GraphQL                        \\
\multicolumn{1}{c|}{T6}         & GraphQL                        & REST                           \\
\multicolumn{1}{c|}{T7}         & REST                           & GraphQL                        \\
\multicolumn{1}{c|}{T8}         & GraphQL                        & REST                           \\ \bottomrule
\end{tabular}
\label{tab:template}
\vspace{18pt}
\end{table}

\subsection{Within-Subject Design}

The treatment in our experiment is the technology to implement the queries, i.e., REST or GraphQL. The dependent variable is the time the subjects take to implement the proposed tasks. We also analyze the results under three dimensions: (i) types of queries (\emph{search repositories}, \emph{search users}, \emph{repository}, and \emph{user}), (ii) students level (undergraduate and graduate), and (iii) previous experience with REST and GraphQL.

The experiment followed a \emph{within-subject design}~\cite{seltman2012}, where all participants were exposed to every treatment. 
In other words, all participants implemented tasks using REST and GraphQL (four tasks in each technology). 
However, it is well-known that the order in which the treatments are given affects the subjects' performance. To counteract this fact, we also used a \emph{counterbalanced design}~\cite{seltman2012}.
Basically, we elaborated two task description documents (A and B) with the tasks alternating between REST and GraphQL, as described in Table~\ref{tab:template}.  
We divided the subjects into two groups, one group received document A, and the other received document B. Table~\ref{tab:groups} shows that both groups are balanced, regarding their programming experience and academic level. However, in terms of experience in REST and GraphQL, Group B has more participants with experience in REST. We could not achieve an uniformed distribution in this case due to last changes in the list of participants (some confirmed participants did not appear and others appeared, despite having not answered our initial invitation). However, this fact did not impact our findings; indeed, as reported in Section~\ref{sec:results}, GraphQL queries were implemented in less time than REST ones.

\begin{table}[h!]
\centering
\caption{Group Profiles}
\begin{tabular}{ccccc}
\toprule
\multirow{2}{*}{\textbf{Group}} & \multicolumn{4}{c}{\textbf{Experience in REST and GraphQL}}                                                           \\ \cmidrule{2-5} 
                                & \textbf{REST}       & \textbf{GraphQL}     & \textbf{REST and GraphQL} & \textbf{None}            \\ \midrule
\multicolumn{1}{c|}{A}          & 4                  & 0                   & 2                        & 5                       \\
\multicolumn{1}{c|}{B}          & 7                  & 0                   & 2                        & 2                       \\ \midrule
\multirow{2}{*}{\textbf{Group}} & \multicolumn{4}{c}{\textbf{General Programming Experience (years)}}                                                   \\ \cmidrule{2-5} 
                                & \multicolumn{2}{c}{\textbf{\textless{}1}}  & \textbf{1..3}             & \textbf{\textgreater{}3} \\ \midrule
\multicolumn{1}{c|}{A}          & \multicolumn{2}{c}{0}                     & 2                        & 9                       \\
\multicolumn{1}{c|}{B}          & \multicolumn{2}{c}{0}                     & 2                        & 9                      \\ \midrule
\multirow{2}{*}{\textbf{Group}} & \multicolumn{4}{c}{\textbf{Academic Level}}                                                        \\ \cmidrule{2-5} 
                                & \multicolumn{2}{c}{\textbf{Undergraduate}} & \multicolumn{2}{c}{\textbf{Graduate}}                \\ \midrule
\multicolumn{1}{c|}{A}          & \multicolumn{2}{c}{5}                     & \multicolumn{2}{c}{6}                               \\
\multicolumn{1}{c|}{B}          & \multicolumn{2}{c}{5}                     & \multicolumn{2}{c}{6}                      \\ \bottomrule
\end{tabular}
\label{tab:groups}

\vspace{7pt}

\end{table}

\subsection{Experimental Procedure}

Before asking the participants to perform the proposed tasks, the first author of this paper presented a short talk (about one hour) on using REST and GraphQL to query GitHub data. After this talk, the subjects completed a pre-experiment form, where they informed their experience with the studied technologies. During the execution of the tasks, the subjects had access to the slides of this initial talk, as well as to the GitHub documentation of both APIs.\footnote{\url{https://developer.github.com/v4/query/}}$^{,}$\footnote{\url{https://developer.github.com/v3/}}

The experiment was conducted using IDLE, which is a simple IDE for programming in Python.\footnote{\url{https://docs.python.org/3/library/idle.html}} IDLE is suitable for beginners, especially in educational environments.
As presented in Listing~\ref{lst-example_task}, we provided to the participants a single source code file, containing the description of the tasks (in the form of comments, see lines 1--8) and specific string variables to store the queries (line 13, for example). After executing the provided code, it automatically informs if the query is correctly implemented or not. If it is correct, we instructed the subjects to move to the next task/query. Otherwise, he/she was instructed to revise and change the implementation and try again. Additionally, each execution generates a log, containing information about the queries (code, result, time, etc). We used this log to compute the time spent on each task $i$, which we called  $T_i$, $1 \leq i \leq 8$. The log also provides the time each task $i$ was concluded ($F_i$). Therefore, $T_i = F_i - F_{i-1}$. All participants started the experiment at the same time, i.e., $F_0$ is known. It is also
important to mention that all participants concluded their eight tasks without interruptions or breaks (e.g., restroom or coffee breaks).

\begin{lstlisting}[caption={Fragment of source code, used to implement Task \#1}, label=lst-example_task]
#======================================================#
# Task 1: Write a query that returns the full name     #
#(owner/name) and the description of the top-10 most   #
#starred Python repositories,                          #
# sorted in descending order.                          #
#  **  First, remove the comment (#) from the used API #
#  **  After, implement the query instead of xxx       #
#======================================================#

#API = 'rest'
#API = 'graphql'

query_1 = """xxx"""
\end{lstlisting}

\begin{figure}[t!]
\centering
%\vspace*{-0.5cm}
\includegraphics[width=.47\textwidth]{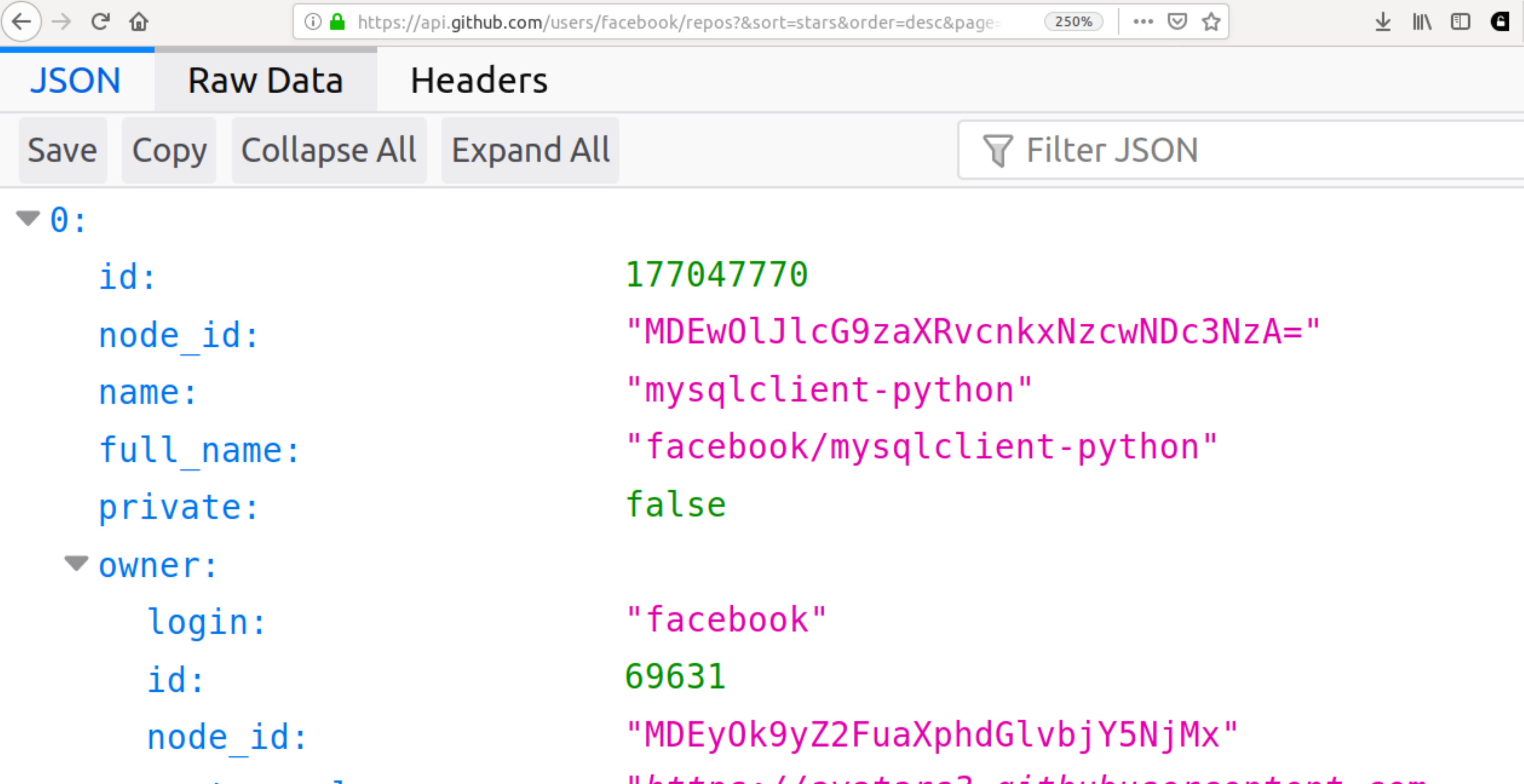}
\caption{Response to a REST query performed using a web browser}

\label{fig:rest_query}
\end{figure}

\begin{figure}[t!]
\centering
\vspace*{0.5cm}
\includegraphics[width=.47\textwidth]{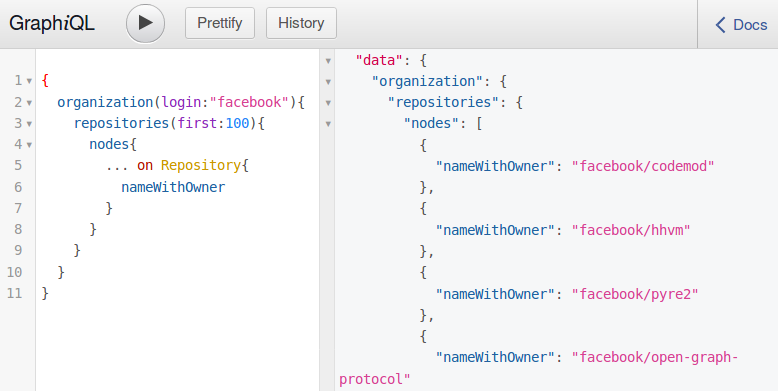}
\caption{Response to a GraphQL query performed using GraphiQL}
\label{fig:graphiQL_query}
\end{figure}

During the experiment, the subjects were allowed to use a web browser to test the queries, particularly the ones implemented in REST. For example, Figure~\ref{fig:rest_query} shows an example of testing a REST query in a web browser. The query is provided in the address bar and the resulting JSON documented is showed in the browser.
Finally, GitHub provides a web app, called GraphiQL, to test GraphQL queries (see an screenshot in Figure~\ref{fig:graphiQL_query}). This app leverages GraphQL features to support for example auto complete. We claim that allowing the participants to use this IDE does not represent a bias towards GraphQL, since it is also used by practitioners in their daily experience with the language (just to reinforce, GraphiQL is an official application, supported by GitHub).

Finally, it is important to mention that all participants concluded the proposed tasks, i.e., no participants had to leave during the experiment or were not able to implement some of the queries.

\section{Results}\label{sec:results}

\noindent \emph{RQ1: How much time do developers spend implementing queries in REST and GraphQL?}
\newline

\begin{figure}[t!]
\centering
%\vspace*{-0.5cm}
\includegraphics[width=0.50\textwidth]{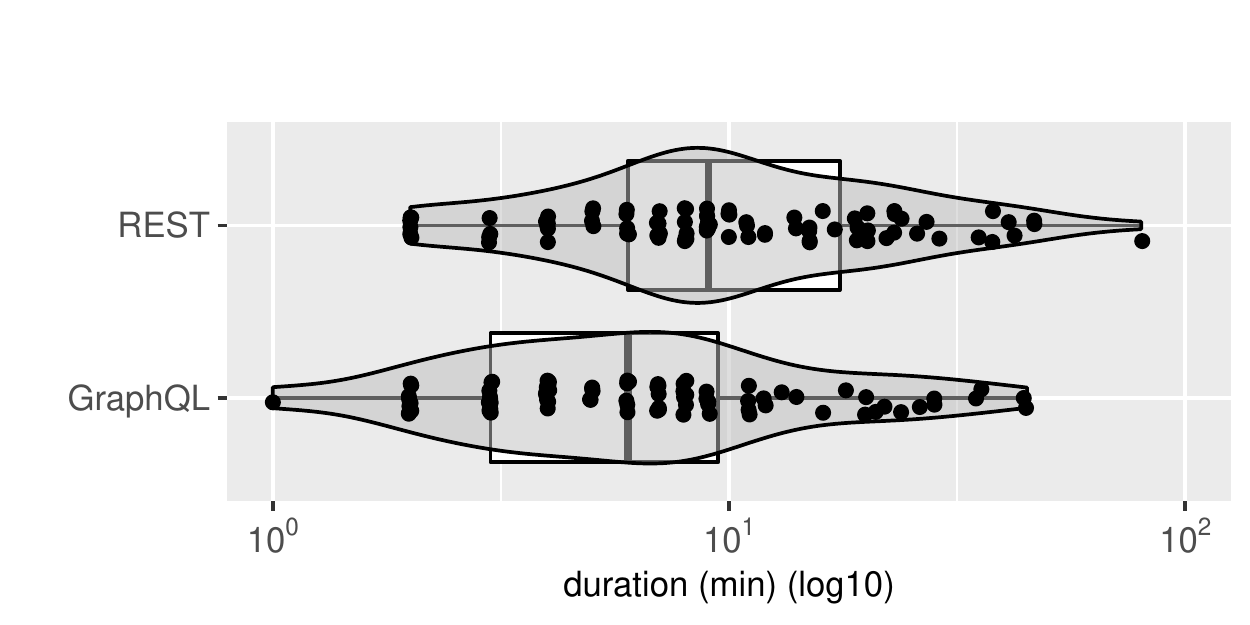}
\caption{Time to conclude the tasks (REST vs GraphQL)}
\label{fig:no_facading}
\end{figure}

Figure~\ref{fig:no_facading} shows violin plots with the time in minutes to implement the proposed tasks using REST and GraphQL. The points in the violins represent the time spent by a \textit{participant} to conclude a \textit{task}. As we can see, the subjects spent on the median nine minutes to implement the REST queries, against six minutes to implement the GraphQL ones (median values). We check these differences by applying a Wilcoxon-Signed Rank test. The \emph{p}-value is 0.00055, which allows us to conclude that the time for implementing the tasks using GraphQL is statistically different than using REST. To show the effect size of the difference between the distributions, we compute  Cliff’s  Delta (\textit{d}-value). Following the guidelines in~\cite{grissom2005, tian2015, linares2013}, the effect size is \textit{medium}.

\vspace{5pt}

\begin{figure*}[t!]
\centering
%\vspace*{-0.5cm}
\includegraphics[width=1\textwidth]{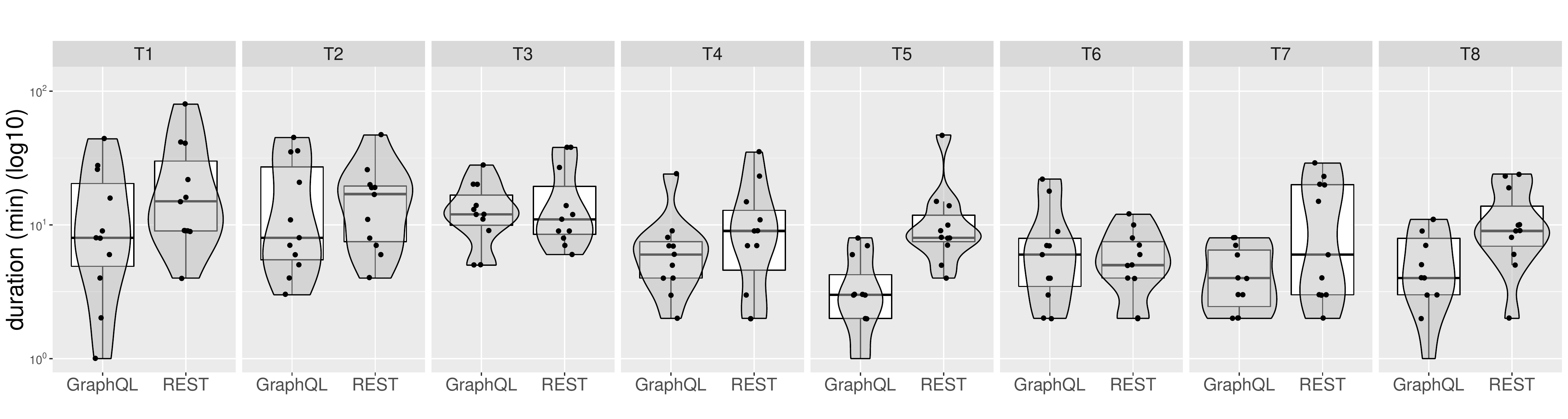}
\caption{Time to implement the proposed tasks for querying GitHub data (REST vs GraphQL)}
\label{fig:facading_task}
\end{figure*}

To shed light on these results, we inspected the performance of the participants in each task.  Figure~\ref{fig:facading_task} shows violin plots with the execution times for each proposed task (as in the previous figure, each point represents the time spent by a \textit{participant} in a given \textit{task}). In this individual analysis, only in task T5 we found a statistical difference
between REST and GraphQL distribution according to a Wilcoxon-Signed Rank test. The effect size of this difference is \textit{large}. Next, we discuss this task in details.

%First we analyze the founded results for T5 and T8, and after we discuss the remaining tasks.  

\vspace{10pt}
\noindent \textit{Task T5.}
This task requires the implementation of a \emph{search users} query to retrieve the top-10 GitHub users with more than 10,000 followers, in decreasing order. 
In this case, the subjects spent eight minutes (median values) to conclude the REST implementation, against only three minutes in GraphQL.

By analyzing the log files of five participants that spent more time than the median when implementing T5 in REST, we found that they all initially implemented the following query: \vspace{1pt}
\begin{lstlisting}
GET /search/users?q=repos:>10000&page=1&per_page=10
\end{lstlisting}
In this query, the \mcode{type} qualifier is missing.
As a result, the query returns data about both personal (e.g., \textit{torvalds}) and organizational (e.g., \textit{facebook}) accounts. The following listing shows the correct REST query, where the \mcode{type} qualifier is added to return information just about \textit{user} accounts. \vspace{7pt}
\begin{lstlisting}[moredelim={[is][keywordstyle]{@@}{@@}}]
GET /search/users?q=repos:>10000+@@type:user@@&page=1&per_page=10 
\end{lstlisting}

\begin{figure}[h!]
\centering
\vspace*{-0.5cm}
\includegraphics[width=0.45\textwidth]{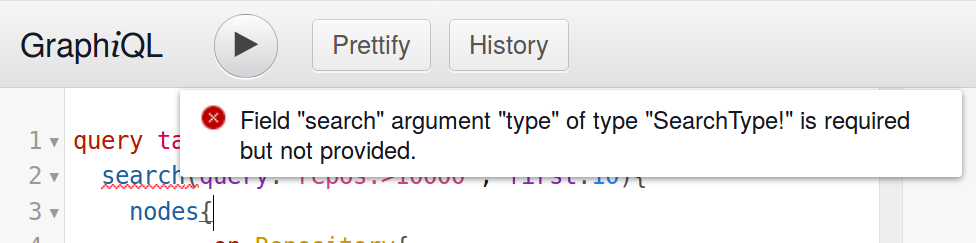}
\caption{Warning message about missing \mcode{type} parameter}
\label{fig:error_type}
\end{figure}

By contrast, for GraphQL, all 11 subjects implemented T5 correctly, using the following query: 
%\vspace{7pt}
\begin{lstlisting}[moredelim={[is][keywordstyle]{@@}{@@}}]
query t5{
    search(query:"repos:>10000", @@type:USER@@, first:10 ){
        nodes{
            ... on User{ login }
        }
    }
}
\end{lstlisting}

As we can see in the previous listing, the \mcode{type} parameter is mandatory in GraphQL (line 2). When this parameter is missing,  GraphiQL (the IDE used by the participants) presents a warning, as showed in Figure~\ref{fig:error_type}. 

\noindent\fcolorbox{lightgray!20}{lightgray!20}{
\begin{minipage}{0.467\textwidth}
\emph{RQ1's summary:} 
When comparing the implementation time of all tasks, there is a difference of 3 minutes favoring GraphQL (9 minutes vs 6 minutes, median times). However, in only one task we found a statistical difference with a \textit{large} effect size. A missing parameter in the REST endpoint was responsible for this difference.
\end{minipage}
}
\vspace*{0.1cm}

\noindent \textit{RQ1.1: How does the implementation time vary among the types of queries?}

%\subsubsection{RQ1.2: How does this time vary between the tasks?}

\begin{figure}[t!]
	\centering
	%\vspace*{-0.5cm}
	\includegraphics[width=0.45\textwidth]{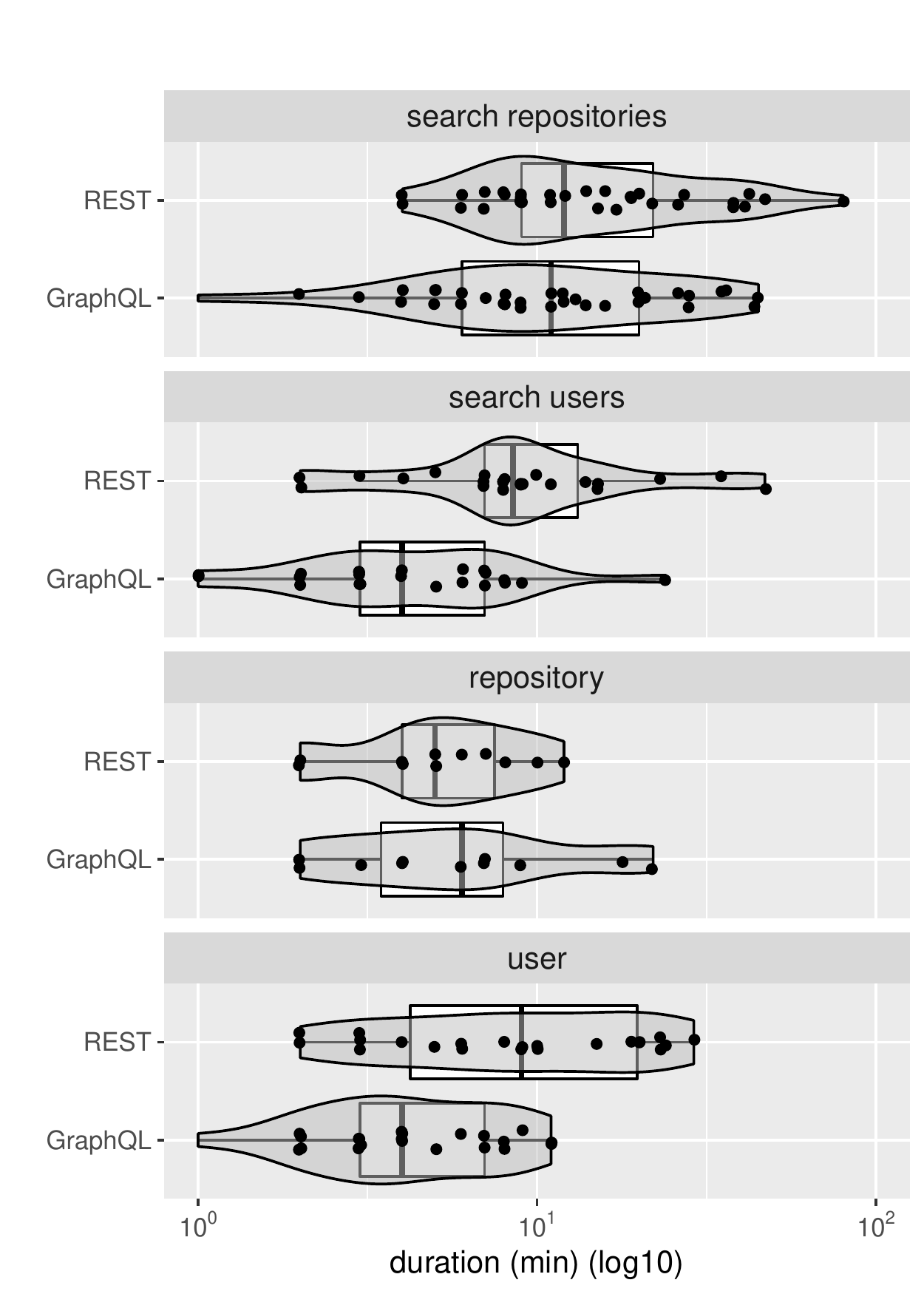}
	\caption{Time to conclude the tasks, grouped by query type}
	\label{fig:facading_type}
\end{figure}

In this RQ, we compare the tasks grouped by query types (\textit{search repositories}, \textit{search users}, \textit{repository}, and \textit{user}). Figure~\ref{fig:facading_type} shows violin plots with the results. As we can see, for three query types (\textit{search repositories}, \textit{search users}, and \textit{user}) the median implementation time was higher when the tasks were implemented in REST. However, according to Wilcoxon-Signed Rank test, only \textit{search users} and \textit{users} present statistical difference. The effect size for \textit{search users} is \textit{large} and for \textit{users} is \textit{medium}.

We hypothesize that queries that return user elements require more effort to be implemented in REST because they demand several parameters. In GraphQL, these parameters are recommended by the auto complete feature of the GraphiQL IDE. For instance, the implementation of T1 by participant S16 nicely illustrates the problems associated to the use of REST parameters. This participant spent 80 minutes to implement T1 (which is 81\% greater than T1's median implementation time in REST). Indeed, S16 spent 125 minutes to conclude all queries. Therefore, only in T1---his first query---he spent 64\% of his overall implementation time. By contrast, the maximum time for implementing T1 in GraphQL was 44 minutes, by S9. This task demands the implementation of a query returning the full name and description of the top-10 most starred Python repositories in descending order. The following listings shows some attempts, performed by S16:
\begin{lstlisting}[numbers=none]
search/repositories?q=language:python+stars&sort=stars&order=desc
\end{lstlisting}
\begin{lstlisting}[numbers=none]
search/repositories?q=language:python+stars:>100&sort=stars&order=desc
\end{lstlisting}
\begin{lstlisting}[numbers=none]
search/repositories?q=language:python+stars:>10&sort=stars&order=desc
\end{lstlisting}
\begin{lstlisting}[numbers=none]
search/repositories?q=language:python+stars&sort=stars&order=desc&page=1&per_page=10
\end{lstlisting}

In the first three queries, S16 did not inform the \mcode{page} and \mcode{per\_page} parameters, which are mandatory parameters for defining the number of returned elements. In the first and fourth queries, he did not inform the value of the \mcode{stars} parameter, which is necessary to select the most starred repositories.

It is interesting to mention that S16 spent 73 minutes to conclude \textit{all} GraphQL tasks (i.e., overall, he spent 63\% of his experiment time in REST and 37\% in GraphQL). The GraphQL task he spent more time was in T2 (45 minutes, which is 82\% greater than T2's median implementation time in GraphQL). We emphasize that S16 had no previous experience with REST or GraphQL. Therefore, tasks T1 and T2 were his first contact with both technologies.

%Additionally, subject S9 has previous experience with REST.

By contrast, only in the case of the \textit{repository} task (T6), we observed higher implementation times for GraphQL, on the median. In this task, the REST endpoint \mcode{/repos/\{owner\}/\{repo\}} is used to implement the proposed task.
This endpoint demands only two parameters, as we can see in the following listing.\\[-0.4cm]
\begin{lstlisting}
GET /repos/facebook/react
\end{lstlisting}
\vspace*{-0.17cm}
On another hand, to implement T6 in GraphQL, besides the two parameters \textit{owner} and \textit{name} (line 2 in the following listing), it is necessary to specify filters to return only the fields mentioned in the specification of T6, i.e., primary language, description, and URL.\\[-0.4cm]
\begin{lstlisting}
{
	repository(owner:"facebook", name:"react"){
		primaryLanguage	{ name }
		description
		url
	}
}
\end{lstlisting}

However, even by REST presenting better results for T6, the difference is of only one minute (5 minutes in REST, against 6 minutes in GraphQL). Indeed, by applying Wilcoxon-Signed Rank test, we did not find a statistical difference in this case.

\vspace{7pt}
\noindent\fcolorbox{lightgray!20}{lightgray!20}{
\begin{minipage}{0.467\textwidth}
\emph{RQ1.1's summary:} 
GraphQL outperforms REST mainly in queries that require several parameters. In such queries, auto complete---as provided by GraphQL's IDEs---is a powerful feature to help developers. For example, a novice developer spent 63\% of his time in REST and 37\% in GraphQL.
\end{minipage}
}
\vspace*{0.1cm}

\noindent \textit{RQ1.2: How does the implementation time vary among undergraduate and graduate students?} \\[0.2pt]

\begin{figure}[t!]
\centering
\vspace*{-0.5cm}
\includegraphics[width=0.47\textwidth]{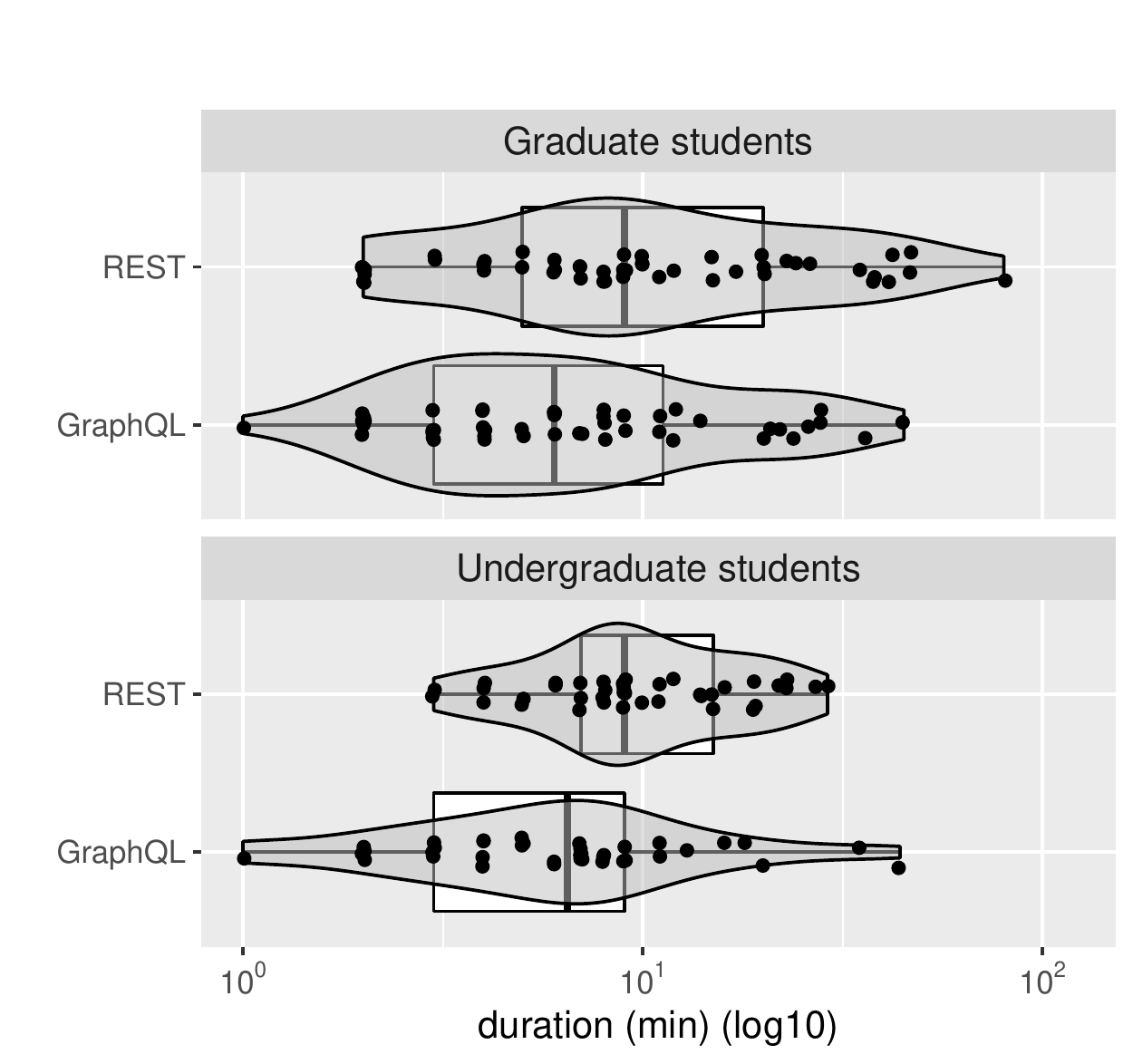}
\caption{Time to conclude the tasks grouped by academic level}
\label{fig:facading_group}
\end{figure}

Figure~\ref{fig:facading_group} presents the results according to the subjects' academic
level. In both groups, the participants who performed the
tasks in REST spent more time than the ones who
implemented them in GraphQL. Therefore, even subjects with more experience, as is typically the case of graduate
students, take benefit of GraphQL. Indeed,
the highest difference between the median times (REST $-$
GraphQL) was observed for graduate students,
3 minutes, against 2.5 minutes
for undergraduate students. According to Wilcoxon-Signed Rank tests,
both distributions (graduate and undergraduate) are statistically different. The effect size is \textit{medium} for both groups.

\vspace{9pt}
\vspace*{0.1cm}
\noindent\fcolorbox{lightgray!20}{lightgray!20}{
\begin{minipage}{0.467\textwidth}
\emph{RQ1.2's summary:} 
Both undergraduate and graduate students
have taken benefit of GraphQL and implemented the tasks in less time.
\end{minipage}
}
\vspace*{0.1cm}

\begin{figure}[t!]
\centering
%\vspace*{-0.5cm}
\includegraphics[width=0.47\textwidth]{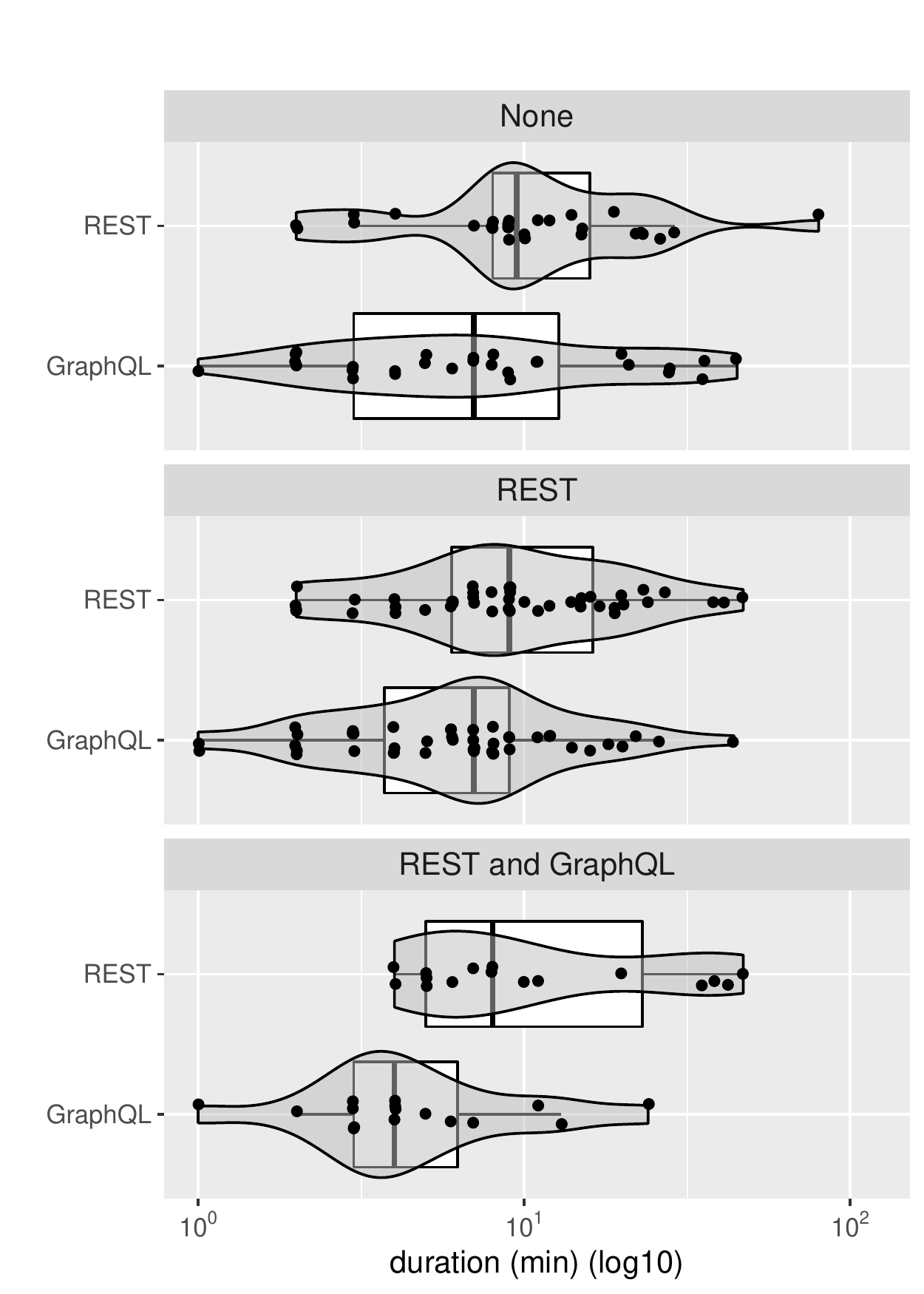}
\caption{Time to conclude the tasks (REST vs GraphQL) grouped by previous experience}
\label{fig:facading_experience}
\end{figure}

\noindent \textit{RQ 1.3 How does the implementation time vary depending on the participants previous experience in REST and GraphQL?} \\[0.2pt]

Figure~\ref{fig:facading_experience} presents the results according to the subjects' previous
experience in REST or GraphQL. Each point in the violin plots
represents the time spent by a participant in the proposed
implementation tasks. As in the previous RQs, we checked the statistical differences
using Wilcoxon-Signed Rank tests. We found a statistical difference
in the last two groups, i.e., participants with
previous experience in REST and participants with 
previous experience in both technologies. The effect size for the REST group is \textit{medium}. In other words, these
participants spent more
time implementing their tasks in REST than
in a completely novel technology for them, i.e., GraphQL.\\[0.2cm]

\noindent\fcolorbox{lightgray!20}{lightgray!20}
{\begin{minipage}{0.467\textwidth}
\emph{RQ1.3's summary:} 
GraphQL outperforms REST even among
participants with previous experience in REST and
no previous contact with GraphQL.
\end{minipage}}
\vspace*{0.2cm}

\vspace{10pt}
\noindent \emph{RQ2: What are the participants' perceptions about REST and GraphQL?}
\newline

After implementing the tasks, the participants were invited to answer a post-experiment survey, with their perceptions about REST, GraphQL, and the experiment. We received responses of 11 participants (3 undergraduate and 8 graduate).

The first author of this paper carefully read these responses and grouped them in four categories: Tool support, Syntax support, Previous Experience, and Documentation. Table~\ref{tab:responseForm} shows the participants with answers in each category. We received at least three answers in all categories, except in Previous Experience.

\begin{table}[h!]
\centering
\caption{Answer's Categories}
\begin{tabular}{l|l}
\toprule
\textbf{Category} & \textbf{Participants} \\ \midrule
Tool Support & S1, S7, S8, S17 \\
Syntax Support  & S9, S12, S17 \\
Previous Experience     & S1    \\
Documentation  & S4, S11, S7      \\ \bottomrule
\end{tabular}
\label{tab:responseForm}
\end{table}

In the following paragraphs, we discuss the answers of each category. 
%\\[0.1pt]

\noindent \textit{Tool Support.} 
Two participants (S1, S17) explicitly mentioned the benefits of using GraphiQL IDE. As examples, we have these answers: \textit{queries with GraphQL are much more interesting to do because of the ease of Explorer (S1)} and \textit{the GraphQL environment helped a lot during query testing (S17)}. 
Furthermore, GraphQL allows IDEs to implement an auto complete feature due to the avaiability of a schema for defining types and fields. This feature was a powerful tool to assist the implementation of queries, as mentioned by subject S7: \textit{the auto complete feature of GraphQL has helped me a lot to put together complex queries}.\\[0.1pt]

\noindent \textit{Syntax Support.}
GraphQL queries follow a JSON syntax. For instance, they can be indented for better understanding and visualization of the code, as mentioned by subject S12: \textit{for me GraphQL is better than REST, because the query structure allows a better visualization of the query}. Furthermore, S17 mentioned that GraphQL is a more intuitive technology because it relies on programming language concepts, like types, functions, and queries: \textit{I think the syntax of the query looks much more like what we are used to see in programming languages (functions, types, SQL queries)}.
Additionally, when using GraphQL, it is easier to define the parameters required by the proposed queries, as mentioned by subject S17: \textit{I liked the GraphQL filters because they allowed me to better specify the characteristics of the query}.\\[0.1pt]

\noindent \textit{Experience.}
Seven participants have no previous experience in GraphQL, but only S1 complained about this fact: \textit{my main difficulty was the lack of previous experience with GraphQL}. Although S1 mentioned his lack of experience, he spent only 41 minutes to implement all GraphQL queries, against 146 minutes for REST. Additionally, he also does not have experience in REST.\\[0.1pt]

\noindent \textit{Documentation.}
During the experiment, the participants had access to REST and GraphQL API documentation. Two participants mentioned that GitHub's GraphQL API documentation is limited, e.g., \textit{API documentation about GitHub's GraphQL is poor (S11)}.\\[0.1pt]

\vspace{7pt}
\vspace*{0.1cm}
\noindent\fcolorbox{lightgray!20}{lightgray!20}{
\begin{minipage}{0.467\textwidth}
\emph{RQ2's summary:} 
According to the subjects, the main benefit of GraphQL are the tool support provided by GraphiQL, e.g., auto complete feature. Another mentioned benefit is a better syntax to read the code and less effort to specify parameters. By contrast, two participants commented about the poor quality of GitHub's GraphQL API documentation.
\end{minipage}
}

\section {Discussion}\label{sec:discution}

\subsection{Why GraphQL requires less effort?}

By triangulating the results of RQ1 and RQ2, it is clear that
the avaliability of a type system---expressed as a schema---
is one of the key benefits provided by GraphQL, in terms
of reducing the effort to implement queries, when compared
to REST.
Essentially, this schema allows GraphQL IDEs to check type
errors before submitting the queries and
also to provide messages with a clear indication
of the errors made by developers. This happened
for example in T5, when all REST participants forgot
to include an important parameter; in GraphQL, the
same error was detected by the GraphiQL IDE, which produced
a clear error message. Consequently, T5 was
implemented in eight minutes in REST against three
minutes in GraphQL (median times).

However, it is also important to clarify that it is GraphQL's schema that allows the implementation of IDEs with features like code completion. These features helped developers in many tasks in our study. Therefore, \textbf{it is unfair to attribute the gains observed with GraphQL only to the IDE}. In fact, the root factor is the language design of GraphQL, which is centered around a type system, by the language designers.

\subsection{Can we improve REST results?}

Interestingly, our results suggest that REST may also benefit from introducing a type system
in endpoints. In other words, for each endpoint,
this type system would describe the required parameters,
their category (mandatory or optional, for example)
and the type of the respective arguments. In this way,
it might be possible to develop REST IDEs with features
similar to the ones of the GraphQL IDE used in our experiment,
including the support to auto complete. Furthermore,
these IDEs could also have a Web browser version,
in order to allow in-browser type checking of REST
queries.

In fact, there are efforts to introduce a type system in REST. For example, OpenAPI Specification\footnote{\url{https://swagger.io/specification/}} is a standard that allows developers to describe the types returned by REST APIs. However, languages such as OpenAPI are not widely used by REST API developers. Therefore, \textbf{our study shows the importance of providing type specifications for REST APIs}.

\section{Threats to Validity}\label{sec:threats}

In this section, we report threats to validity, as well as, the respective treatments, based on the guidelines proposed by Wohlin et al.~\cite{wohlin2012}.

%\subsection{Construct Validity}

\vspace{5pt}
\noindent \textit{Construct Validity.}
The construct validity in controlled studies refers to correctly measuring of the dependent variables, which in our case, is the time to implement the proposed queries. A possible threat to the experimental procedure is the possibility of dialogue between the experimenter and the subjects, interfering in the process to implement the proposed queries.  Therefore, the first author of this paper carefully asked the students to avoid discussing and commenting their work during the experiment.

\vspace{5pt}
\noindent \textit{Internal Validity.}
The  internal  validity  is  related  to  uncontrolled aspects that may affect the experimental results, since the subjects' experience. To mitigate this threat, we distributed the subjects in two groups with the tasks alternating between REST and GraphQL (\textit{counterbalanced design}). We also equally distributed graduate and undergraduate students in these two groups. Another threat is the use of statistical machinery. We  paid  special  attention  to  the  appropriate  use  of  statistical tests (i.e., Wilcoxon-Signed Rank) when reporting our results in RQ1.  This  reduces  the  possibility  that our findings are due to random events.

\vspace{5pt}
\noindent \textit{External Validity.}
The external validity is related to the possibility to generalize our results. The experiment was conducted with 22 subjects. Thus, this number of subjects might not be a representative sample. However, our sample is diversified; the subjects have different academic levels, general programming experience, and previous experience with REST and GraphQL. Additionally, the number of proposed tasks is another possible threat. In our study, we used eight tasks to measure the effort to implement REST and GraphQL queries. However, we investigate four different types of queries, with different difficult levels. The proposed queries were prepared by the authors based in real queries used in empirical software engineering papers. Moreover, we compare REST and GraphQL using a single API (GitHub API). However, we are not aware of other public and large API, that support both REST and GraphQL. A final threat is the fact that our subjects are students. However, according to previous studies~\cite{salman2015, host2000, runeson2003}, students may provide an adequate model of professional developers.

\section{Related Work}\label{sec:relWork}

We separated related work in four categories: (a) studies about controlled experiments; (b) studies on REST and SOAP technologies; (c) studies on the query language GraphQL; and (d) studies on other programming languages.

\subsection{Controlled Experiments}

Controlled experiments have been widely adopted in software engineering research as a way to evaluate tools and technologies~\cite{wohlin2012}. 
According Wohlin et al.~\cite{wohlin2012}, controlled study is an empirical strategy that manipulates one factor of the studied setting where different treatments are applied to one or more variables, while other variables  are kept constant.
Avidan et al.~\cite{avidan2017} conducted a controlled experiment where nine developers tried to understand six methods from utility classes, either with the original variable names or with names replaced by meaningless single letters. This study shows that parameter names are more significant for comprehension than local variables. Melo et al.~\cite{melo2016}, perform a controlled experiment to quantify the impact of variability on debugging of preprocessor-based programs. They measured the speed and precision for bug finding tasks at three different degrees of variability on several real systems. As well in these previous studies, we also apply a controlled experiment to quantify the impact of REST and GraphQL adoption in the time to implement queries.

\subsection{REST and SOAP Studies}

Two consolidated technologies for Web Services design are REST and SOAP (Simple Object Access Protocol).
AlShahwan et al.~\cite{alshahwan2010}, they perform a comparison between frameworks to implement  SOAP and REST services with focus on such devices. They conclude that REST is more suitable for mobile environments. 
Mulligan et al.~\cite{mulligan2009} assess the effectiveness of SOAP and REST in satisfying key backend data transmission requirements. To this purpose, the authors provide implementations of a data transmission service using SOAP and REST. Finally, they evaluate both implementations with emphasis on performance, efficiency, and scalability. They conclude that REST is more efficient for data transmission.
There are also studies on migrating SOAP services to REST. Upadhyaya et al.~\cite{upadhyaya2011} identify resources from a SOAP web service by analyzing the service description and migrating each service to a REST architecture. 
Their approach consists on the identification of similar operations, resources, and methods. They also conduct a case study to evaluate the approach. As a result, the authors conclude that the performed migration improves the performance of the migrated services in 74\%.

\subsection{GraphQL Studies}

Since it is a recent technology, there are few studies in the scientific literature on GraphQL. Hartig and Perez~\cite{hartig2017} provide a formal definition for GraphQL. Recently, the authors complemented and finished this formalization by proving that evaluating the complexity of GraphQL queries is a NL-problem~\cite{hartig2018}. In practical terms, this result shows that it is possible to implement efficient algorithms to estimate the complexity of GraphQL queries before their execution; which is important to handle the performance problems normally associated to GraphQL. Vogel et al.~\cite{vogel2017} present a case study on migrating to GraphQL part of the API provided by a smart home management system. They report the runtime performance of two endpoints after migration to GraphQL. For the first endpoint, the gain was not relevant; but for the second, GraphQL required 46\% of the time of the original REST API.
Wittern et al.~\cite{wittern2018} assess the feasibility of automatically generating GraphQL wrappers for existing REST(-like) APIs. For this purpose the authors propose a tool to generate GraphQL wrappers from REST-like APIs with OpenAPI Specification (OAS). Their tool takes as input a specification that describes a REST API and automatically generates a GraphQL wrapper. The proposed tool was evaluated with 959 publicly available REST APIs and was able to generate GraphQL wrappers for 89.5\% of these APIs, with limitations in some cases. Wittern et al.~\cite{wittern2019} also perform a study on GraphQL schemas. The authors study the design of GraphQL interfaces by analyzing schemas of 8,399 GitHub projects and 16 commercial projects. The authors report that a majority of GraphQL APIs have complex queries, posing real security risks. Vargas et al.~\cite{vargas2018} perform a study to investigate the feasibility of using a classic technique to test generation in GraphQL schemas (deviation testing). They use an implementation of GraphQL for Pharo and run the proposed technique in two popular GraphQL APIs. 
Finally, Brito et al.~\cite{brito2019} perform a study on migrating GitHub clients from REST to GraphQL API. First, the authors conduct a grey literature review to understand the characteristics and benefits of GraphQL adoption. After, they assess these benefits by migrating seven systems to use GraphQL instead of REST APIs. They conclude that GraphQL can reduce the size of the JSON documents in 99\% (number of bytes).

\subsection{Programming Language Studies}
There are also studies investigating the impact of programming languages in software quality and development time. For example, Ray et al.~\cite{ray2014} investigated the impact of programming languages on software quality. For this purpose, the authors perform a study with 729 GitHub systems. The results point that strong typing is slightly better than weak typing, and functional languages are somewhat better than procedural languages. Another study on impact of programming languages on code quality was conduced by Bhattacharya and Neamtiu~\cite{bhattacharya2011}. In this study, the authors investigate how the choice of programming language impacts software quality. They conduct a study and statistical analysis on four popular open source projects. The authors only consider projects that have considerable portions of development in C and C++. The main finding is that by using C++ instead of C results in improved software quality and reduced maintenance effort.

\section{Conclusion}\label{sec:conclusion}

This paper presented a controlled experiment to investigate the effort to implement API queries in REST and GraphQL. As our key finding, we found that GraphQL requires less effort to implement API queries, when compared with REST. We also showed that (i) queries whith many parameters are particularly more difficult to implement in REST than in GraphQL; (ii) we also observe that GraphQL requires less effort even for developers that have no previous experience which this technology. Also, interestingly, experts in REST APIs can also write GraphQL queries with less effort. In our study, we also investigate the perceptions of the subjects. Most of them related that the main benefit of GraphQL is the tool support provided by GraphiQL, e.g., auto complete feature. Another benefit is a better syntax to understanding the code and less effort to specify parameters.
As future work, we intend to extend this research by surveying and interviewing practitioners to reveal their views and experience with GraphQL.
We also intend to investigate possible challenges to adopt GraphQL in real systems, e.g., by migrating REST APIs to GraphQL. Another future work is to investigate the development of REST
IDEs  with  features similar to the  ones of GraphQL
IDEs.

\section*{Acknowledgments}

We deeply thanks the 22 students who participated in the experiment and therefore helped us reaching the findings exposed in this paper. Other acknowledgments are omitted due to DBR policy. .

\bibliographystyle{IEEEtran}
\balance

\bibliography{bibtex}
\end{document}